\documentclass[conference]{IEEEtran}
\IEEEoverridecommandlockouts
\usepackage{cite}
\usepackage{amsmath,amssymb,amsfonts}
\usepackage{algorithm}
\usepackage{algpseudocode}
\usepackage{graphicx}
\usepackage{textcomp}
\usepackage{xcolor}
\usepackage{optidef}
\usepackage{subcaption}
\usepackage{placeins}

\usepackage{stfloats}

\AtBeginEnvironment{subequations}{\vspace{-2pt}}
\AtEndEnvironment{subequations}{\vspace{-2pt}}

\def\BibTeX{{\rm B\kern-.05em{\sc i\kern-.025em b}\kern-.08em
    T\kern-.1667em\lower.7ex\hbox{E}\kern-.125emX}}
\begin{document}

\title{\huge Joint Energy Efficiency Optimization for Uplink Multiuser Movable Antenna-Based Wireless Systems Assisted by Movable-Element RIS\\
}


\author{
\IEEEauthorblockN{Ayda Nodel Hokmabadi, Mohamed Elhattab, Chadi Assi}
\IEEEauthorblockA{
Concordia University, Montreal, Canada\\
a\_nodelh@encs.concordia.ca, mohamed.elhattab@concordia.ca, chadi.assi@concordia.ca}
}

\maketitle

\begin{abstract}
This paper investigates energy efficiency (EE) optimization for an uplink multiuser system assisted by a movable-element reconfigurable intelligent surface (ME-RIS) and a base station equipped with movable antennas (MA-BS). We jointly optimize the uplink postcoder vectors, 
user transmit powers, RIS phase shift, and the positions of both the BS antennas and 
RIS elements to maximize the system EE. The resulting non-convex fractional problem is solved 
using an alternating optimization (AO) framework, where subproblems are handled via 
Dinkelbach's method combined with successive convex approximation (SCA). Simulation results 
show that the proposed scheme achieves significant EE gains over fixed-antenna BS and fixed-element RIS benchmarks.
\end{abstract}

\begin{IEEEkeywords}
Energy efficiency, movable antennas, movable-element reconfigurable intelligent surface, uplink, 6G.
\end{IEEEkeywords}

\section{Introduction}
Reconfigurable intelligent surface (RIS) has emerged as a promising technology for enhancing spectral and energy efficiency in 6G wireless networks~\cite{basar2019wireless}. By passively reflecting incident signals with controllable phase shifts, RIS can reshape the wireless propagation environment without requiring active radio-frequency chains, making it an energy-efficient solution for next-generation communication systems~\cite{huang2019ris, wu2019irs}. More recently, movable antenna (MA) systems have been proposed to utilize spatial degrees of freedom. Unlike conventional fixed-position antennas, MAs can be repositioned within a limited region, allowing the system to improve channel conditions by adjusting antenna locations~\cite{zhu2024ma_mag, zhu2024ma_twc}. This additional flexibility has been shown to provide significant gains in beamforming performance and interference management.

However, movable antennas and conventional RIS still have inherent limitations when used individually. Conventional RIS can only adjust phase shifts with fixed element locations, limiting adaptation to spatial channel variations. Similarly, movable antennas can mitigate fading or blockage but cannot exploit programmable reflections. These limitations motivate movable-element RIS (ME-RIS), where each reflecting element can adjust both its physical position within a limited region and its phase shift~\cite{zhao2025me_ris, zhang2024ma_ris_geometry}. Such dual reconfigurability enables more flexible channel shaping and improved performance~\cite{wei2024movable_ris, li2024mis}. By jointly optimizing element positions and phase shifts, ME-RIS provides additional spatial degrees of freedom to shape the wireless environment, improving channel gains and interference management. Prior works have studied ME-RIS for downlink sum-rate maximization in multiuser MISO and single-user SISO systems, typically assuming fixed-position BS antennas (FPA-BS)~\cite{MERIS2, MERIS3}. Secure communication for ME-RIS-aided systems with FPA-BS has also been studied through sum secrecy rate maximization~\cite{MERIS1}.

Motivated by these technologies, several works have studied the joint use of RIS and movable antennas~\cite{zhang2025ris_ma, wei2024ma_irs, hokmabadi2026joint, zhao2026exploiting}. However, most existing studies mainly focus on spectral efficiency and assume either fixed RIS element positions or fixed base station (BS) antenna locations, and none have considered the joint design of ME-RIS with movable-antenna base stations (MA-BS) under an energy efficiency (EE) objective. Such assumptions limit the available spatial flexibility and prevent the system from fully exploiting the potential performance gains offered by movable architectures. It is worth mentioning that EE has become an important design objective in future wireless networks, particularly in uplink scenarios where user devices operate under tight power budgets. Improving system performance without significantly increasing power consumption is therefore essential. 

Integrating ME-RIS with MA-BS provides additional spatial adaptability. While ME-RIS reshapes the propagation environment through jointly optimized element positions and controllable reflections, movable BS antennas adjust the receive array geometry to better receive the incoming signals. This joint spatial adaptation can improve interference management and signal alignment in multiuser uplink systems, ultimately leading to improved energy efficiency.

\subsection{Contributions}
We consider a multiuser uplink system, where both the BS antennas and the RIS elements are movable, forming an ME-RIS-aided MA-BS architecture. We formulate 
an EE maximization problem that jointly optimizes the receive postcoding vectors at BS, user transmit powers, RIS phase shifts, and antenna/element positions. The non-convex fractional objective 
is tackled through an alternating optimization (AO) framework combining Dinkelbach's method and SCA. The main 
contributions of this work are summarized as follows:
\begin{itemize}
    \item We propose a joint optimization framework for EE maximization in an uplink multiuser system with ME-RIS and MA-BS, subject to per-user quality of service (QoS) constraints and physical placement limits.
    \item We develop an efficient AO-based algorithm where each subproblem is solved with 
    guaranteed convergence using Dinkelbach and SCA methods.
    \item Results confirm our proposed scheme consistently outperforms 
    fixed-position benchmarks in energy efficiency.
\end{itemize} 
%
%

\section{System Model} \label{sec:system}
\subsection{Network Model}
\begin{figure}[!t]
\centerline{\includegraphics[width=0.9\linewidth]{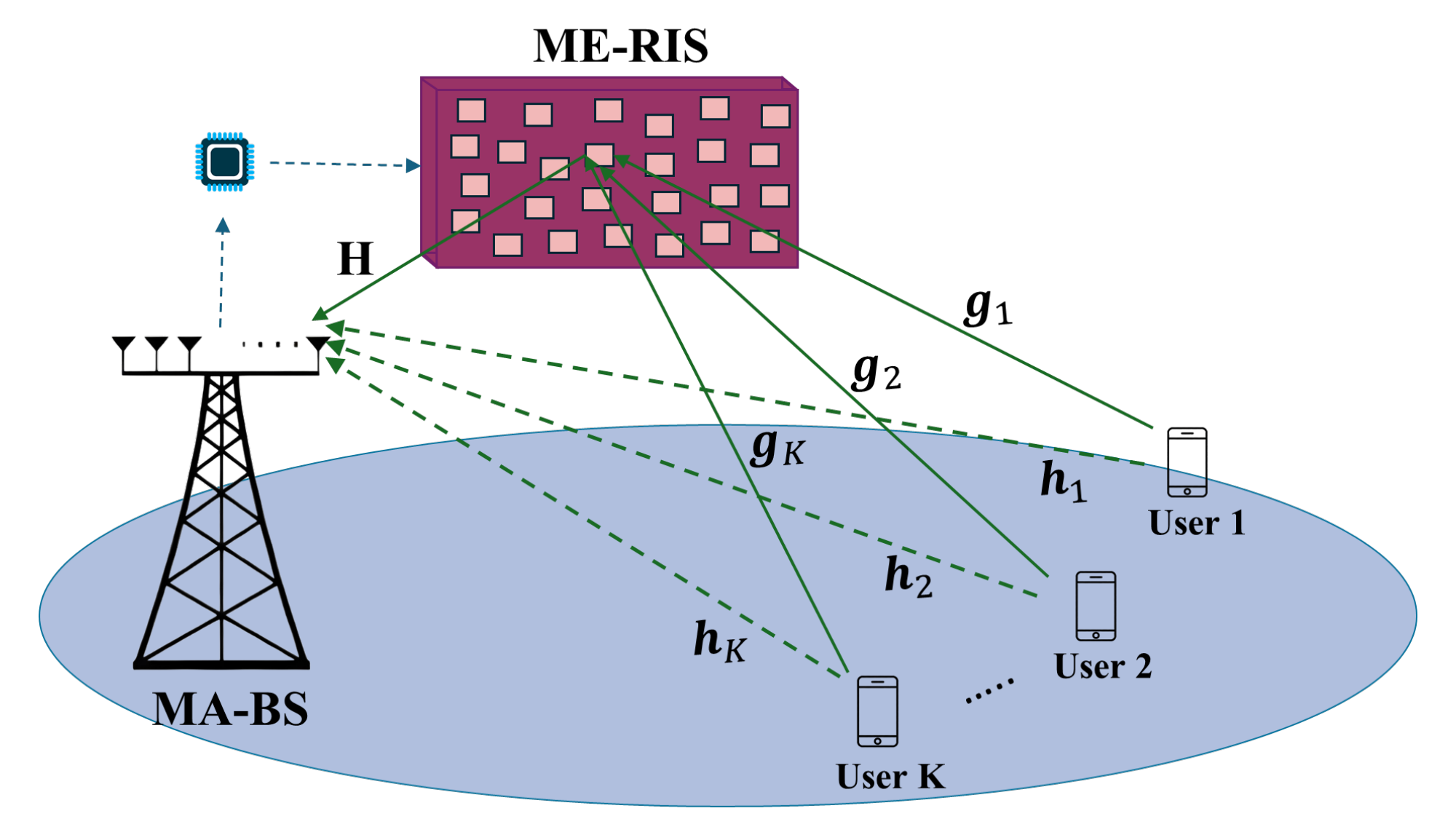}}
\caption{System model}
\label{system_model}
\end{figure}

As shown in Fig.~\ref{system_model}, we consider an uplink single-input multiple-output (SIMO) network consisting of a BS equipped with $M$ movable receiving antennas, a ME-RIS with $N$ passive reflecting elements, and $K$ single-antenna uplink users. The sets of BS antennas, RIS elements and users are denoted by $\mathcal{M}$, $\mathcal{N}$, and $\mathcal{K}$, respectively. 

We denote the channel between the BS antennas and the RIS elements by $\mathbf{H} \in \mathbb{C}^{M \times N}$, the direct link between the user $k$ and the BS by $\mathbf{h}_k \in \mathbb{C}^{M \times 1}, \forall k \in \mathcal{K}$, and the channel between the user $k$ and the RIS by $\mathbf{g}_k \in \mathbb{C}^{N \times 1}, \forall k \in \mathcal{K}$. The noise at the BS is modeled as circularly symmetric complex Gaussian, denoted by $n \sim \mathcal{CN}(0, \sigma^2)$.
\subsection{Signal Model}
Let $s_k$ denote the transmitted  information symbol of user $k$, with $\mathbb{E}\{|s_k|^2\} = 1, \forall k \in \mathcal{K}$. The BS applies a linear postcoder vector $\boldsymbol{v}_k \in \mathbb{C}^{M \times 1}$ to detect the signal of user $k$, and let the transmit power of user $k$ be by $p_k \in [0, P_{\text{max}}]$. Furthermore, the RIS phase shift matrix is defined as a diagonal matrix $\boldsymbol{\Phi} = \text{diag}(e^{j\vartheta_1}, \ldots, e^{j\vartheta_N})$, where $\vartheta_n \in (0, 2\pi], \forall n \in \mathcal{N}$.

Based on that, the received signal at the BS due to the uplink transmission can be expressed as follows,
\enlargethispage{-\baselineskip}
\begin{equation}
\mathbf{y} = \sum_{k \in \mathcal{K}}{\sqrt{p_k} \left(\mathbf{h}_k + \mathbf{H}\boldsymbol{\Phi}\mathbf{g}_k\right) s_k} + \mathbf{n}. \label{eq:sinr_ul}
\end{equation}
After linear combining, the detected signal of user $k$ is $\hat{s}_k = \boldsymbol{v}_k^H \mathbf{y}$. Thus, the signal-to-interference-plus-noise ratio (SINR) $\forall k \in \mathcal{K}$ is
\begin{equation}
\gamma_{k} =
\frac{p_k \left|\boldsymbol{v}_k^{H}\!\left(\mathbf{h}_k + \mathbf{H}\boldsymbol{\Phi}\mathbf{g}_k\right)\right|^2}
{\sum_{\substack{j \in \mathcal{K}\\ j \neq k}}
p_j \left|\boldsymbol{v}_k^{H}\!\left(\mathbf{h}_j + \mathbf{H}\boldsymbol{\Phi}\mathbf{g}_j\right)\right|^2
+ \sigma^2 \|\boldsymbol{v}_k\|^2 } . \label{SINR}
\end{equation}

\subsection{Field-Response Based Channel Model}

Wireless links are modeled using the field-response (FR) geometric channel model~\cite{amhaz2026meta, hokmabadi2026joint}. Since the movable regions of the MA-BS antennas and ME-RIS elements are small relative to link distances, the far-field assumption holds. Hence, the path angles and amplitudes remain approximately unchanged within the movement region, while the phases vary with the antenna and RIS elements positions, enabling channel reconfiguration through position optimization. For a link with $L$ dominant paths, the channel matrix is modeled as
\begin{equation}
\mathbf{C} = \mathbf{F}^H \boldsymbol{\Sigma} \mathbf{E},
\end{equation}
where $\mathbf{E}$ and $\mathbf{F}$ are the transmit and receive field-response matrices, and $\boldsymbol{\Sigma}=\mathrm{diag}(\zeta_1,\ldots,\zeta_L)$ is the path-response matrix, with
$\zeta_\ell \sim \mathcal{CN}\!\left(0,\frac{\beta_0 d^{-\alpha}}{L}\right), \forall l \in\{1,\ldots,L\}.$
Detailed derivation of the FR channel model can be found in~\cite{hokmabadi2026joint}.

\subsubsection{System Channels}

Let $L_{\mathrm{RB}}$, $L_{\mathrm{Bu}}$, and $L_{\mathrm{Ru}}$ denote the number of paths for the RIS--BS, user--BS, and user--RIS links, respectively. The MA-BS antenna positions are $\mathbf{u}_m=[x_m,y_m]^T$ with $\mathbf{U}=[\mathbf{u}_1,\ldots,\mathbf{u}_M]$, and the RIS element positions are collected in $\mathbf{T}$.

The RIS--BS channel is
\begin{equation}
\mathbf{H}=
\mathbf{F}_{\mathrm{RB}}^{H}
\boldsymbol{\Sigma}_{\mathrm{RB}}
\mathbf{E}_{\mathrm{RB}}
\in \mathbb{C}^{M \times N},
\end{equation}
where $\mathbf{F}_{\mathrm{RB}}$ and $\mathbf{E}_{\mathrm{RB}}$ depend on the BS antenna and RIS element positions, respectively. The direct user--BS channel for user $k$ and the user--RIS channel are
\begin{equation}
\mathbf{h}_k =
\mathbf{F}_{\mathrm{Bu},k}^{H}
\boldsymbol{\sigma}_{\mathrm{Bu},k},
\quad
\mathbf{g}_k =
\mathbf{E}_{\mathrm{Ru},k}^{H}
\boldsymbol{\sigma}_{\mathrm{Ru},k},
\end{equation}
where $\boldsymbol{\sigma}_{xy} \in \mathbb{C}^{L \times 1}$ and denotes the path-response vectors in link $xy$ with elements following $\mathcal{CN}\!\left(0,\frac{\beta_0 d^{-\alpha}}{L}\right)$.

\subsubsection{Energy Efficiency}
According to \eqref{SINR}, the network sum rate is given by
\begin{equation}
R_{\text{sum}} = \sum_{k \in \mathcal{K}} \log_2(1+\gamma_k).
\end{equation}
The total power consumption of the network is expressed as
\begin{equation}
P_{\text{tot}} = \frac{1}{\eta}\sum_{k \in \mathcal{K}} p_k + P_c,
\end{equation}
where $\eta \in (0,1]$ denotes the power amplifier efficiency and $P_c$ represents the circuit power consumption.
Therefore, the EE of the system is defined as
\begin{equation}
\mathrm{EE} = \frac{R_{\text{sum}}}{P_{\text{tot}}}.
\end{equation}

\section{Problem Formulation} \label{sec:problem}

We aim to maximize the EE of the multiuser uplink system by jointly optimizing the receive beamforming vectors, RIS phase shifts, user transmit powers, and the positions of the movable antennas at the MA-BS and elements at the ME-RIS. The design is subject to QoS constraints for each user's rate, transmit power limits, unit-modulus constraints on RIS elements, feasible movement regions for antennas/elements, and minimum distance requirements between them. The resulting optimization problem is formulated in problem~\ref{P_EE}.

Constraint~\eqref{C1} guarantees the minimum rate requirement for each user, while \eqref{C2} limits the transmit power. Constraint~\eqref{C3} normalizes the BS receive beamformers and \eqref{C4} enforces the unit-modulus RIS phase shifts. Constraints~\eqref{C5} and \eqref{C6} restrict antenna and RIS element positions within their feasible regions, and \eqref{C7}--\eqref{C8} ensure a minimum spacing $d_0$ to avoid physical overlap.

\begingroup
\setlength{\jot}{2pt}
\interdisplaylinepenalty=10000
\begin{maxi!}|s|[2]
{ \substack{ \mathbf{p}, \boldsymbol{v}_k,\boldsymbol{\Phi}, \mathbf{U}, \mathbf{T} } }
{\mathrm{EE}}
{\label{P_EE}}
{}
\addConstraint{R_k}{\ge R_{\mathrm{th}}, \quad \forall k \in \mathcal{K} \label{C1}}
\addConstraint{0 \le p_k}{\le P_k^{\max}, \quad \forall k \in \mathcal{K} \label{C2}}
\addConstraint{\|\boldsymbol{v}_k\|_2^2}{= 1, \quad \forall k \in \mathcal{K} \label{C3}}
\addConstraint{|[\boldsymbol{\Phi}]_{n,n}|}{= 1, \quad \forall n \in \mathcal{N} \label{C4}}
\addConstraint{\mathbf{u}_m}{\in \mathcal{U}_m, \quad \forall m \in \mathcal{M} \label{C5}}
\addConstraint{\mathbf{T}_n}{\in \mathcal{T}_n, \quad \forall n \in \mathcal{N} \label{C6}}
\addConstraint{\|\mathbf{u}_m - \mathbf{u}_{m'}\|_2}{\ge d_0, \quad \forall m \neq m' \label{C7}}
\addConstraint{\|\mathbf{T}_n - \mathbf{T}_{n'}\|_2}{\ge d_0, \quad \forall n \neq n' \label{C8}}
\end{maxi!}
\endgroup
\vspace{-3pt}
\section{Proposed AO-based algorithms}\label{sec:algorithm}
The main problem is non-convex due to the coupling between variables, the fractional objective function, and the non-convex constraints. To handle this, we decompose it into five subproblems and solve them iteratively. Power allocation is done using the Dinkelbach method combined with SCA, while a closed-form solution is obtained for the uplink postcoder vectors. The BS antenna positions, ME-RIS element locations, and phase shifts are then optimized using first-order SCA. The details of each subproblem are discussed below.
\subsection{Uplink Postcoding Vectors at BS ($\boldsymbol{v}_k$)}
We optimize the uplink receive postcoding vector at the BS for each user $k$, denoted by $\boldsymbol{v}_k$, while keeping $\{\mathbf{p}, \boldsymbol{\Phi}, \mathbf{T}, \mathbf{U}\}$ fixed. The goal is to maximize the sum rate, since the denominator of the EE expression is independent of $\boldsymbol{v}_k$. The corresponding subproblem is
\begin{subequations}
\label{eq:sub_v_original}
\begin{align}
\textup{(P1.1):} \quad
& \max_{\{\boldsymbol{v}_k\}} \quad R_{\mathrm{sum}} \nonumber \\
\mathrm{s.t.} \quad
& R_k \ge R_{\mathrm{th}}, \quad \forall k \in \mathcal{K}, \label{C1_v} \\
& \|\boldsymbol{v}_k\|_2^2 = 1, \quad \forall k \in \mathcal{K}. \label{C2_v}
\end{align}
\end{subequations}

Since $\boldsymbol{v}_k$ appears only in the SINR of user $k$ and is independent of the other users' combining vectors, the subproblem decouples into $K$ independent ones that can be solved in parallel. Each subproblem has a generalized Rayleigh quotient structure~\cite{rayleigh1}. The effective channel for user $k$ is defined as
\begin{align}
\boldsymbol{a}_k \triangleq \mathbf{h}_k + \mathbf{H}\boldsymbol{\Phi}\mathbf{g}_k.
\end{align}
Defining $\mathbf{A}_k \triangleq \boldsymbol{a}_k\boldsymbol{a}_k^H$ and
$\mathbf{B}_k \triangleq \sum_{j \neq k}\boldsymbol{a}_j\boldsymbol{a}_j^H$,
the SINR of user $k$ can be written using the generalized Rayleigh quotient:
\begin{equation}
\gamma_{k} =
\frac{p_k\,\boldsymbol{v}_k^H \mathbf{A}_k \boldsymbol{v}_k}
{\boldsymbol{v}_k^H(\mathbf{B}_k + \sigma^2\mathbf{I}_M)\boldsymbol{v}_k}.
\end{equation}
The optimal combining vector is then given by the dominant generalized eigenvector~\cite{rayleigh3}:
\begin{equation}\label{eq:v_star}
\boldsymbol{v}_k^{\star} =
\frac{(\mathbf{B}_k + \sigma^2 \mathbf{I}_M)^{-1}\boldsymbol{a}_k}
{\left\|(\mathbf{B}_k + \sigma^2 \mathbf{I}_M)^{-1}\boldsymbol{a}_k\right\|_2}.
\end{equation}
After computing $\boldsymbol{v}_k^{\star}$, we check whether the QoS constraint \eqref{C1_v} is satisfied. If it is violated, $\boldsymbol{v}_k^{\star}$ is reset to the previous feasible solution and the AO algorithm proceeds to update the remaining optimization variables.

\subsection{Power Allocation ($\mathbf{p}$)}
Furthermore, we optimize the uplink transmit power vector 
$\mathbf{p} = [p_1, \dots, p_K]^T$ for fixed $\{\boldsymbol{v}_k, \boldsymbol{\Phi}, \mathbf{T}, \mathbf{U}\}$, to maximize the system EE, while satisfying QoS and power constraints.

By defining $A_{k,j} = \left| \boldsymbol{v}_k^H (\mathbf{h}_j + \mathbf{H}\boldsymbol{\Phi}\mathbf{g}_j) \right|^2,$ the achievable rate of user $k$ is given as,
\begin{equation}
R_k(\mathbf{p}) =
\log_2 \left(
1 + \frac{p_k A_{k,k}}
{\sum_{j \neq k} p_j A_{k,j} +\sigma_u^2}
\right).
\end{equation}

The EE maximization problem is therefore,

\begin{subequations}
\label{eq:sub_p_original}
\begin{align}
\textup{(P1.2):} \quad
& \max_{\mathbf{p}} \quad 
\frac{\sum_{k=1}^{K} R_k(\mathbf{p})}
{\frac{1}{\eta} \sum_{k=1}^{K} p_k + P_c}
\nonumber \\
\mathrm{s.t.} \quad
& 0 \le p_k \le P_k^{\max}, 
\quad \forall k \in \mathcal{K}, 
\label{C1_p} \\
& R_k(\mathbf{p}) \ge R_{\mathrm{th}}, 
\quad \forall k \in \mathcal{K}. 
\label{C2_p}
\end{align}
\end{subequations}

The above problem is fractional and non-convex. 
We first apply Dinkelbach’s method~\cite{dinkelbach}, which converts the EE maximization into a sequence of subtractive problems,
\begin{equation}
\max_{\mathbf{p}}
\quad
R_{\text{sum}} - \lambda ({\frac{1}{\eta} \sum_{k=1}^{K} p_k + P_c}),
\end{equation}
where $\lambda$ is iteratively updated as
\begin{equation}
\lambda^\star =
\frac{R_{\text{sum}}}
{\frac{1}{\eta} \sum_{k=1}^{K} p_k + P_c}.
\end{equation}

For a fixed $\lambda$, the objective can be written in difference-of-convex (DC) form. 
Specifically, each user rate can be expressed as
\begin{equation}
\resizebox{0.98\columnwidth}{!}{$
R_k(\mathbf{p})
=
\log_2 \left(
\sigma_u^2 + \sum_{j=1}^{K} p_j A_{k,j}
\right)
-
\log_2 \left(
\sigma_u^2 + \sum_{j \neq k} p_j A_{k,j}
\right)
$}
\end{equation}

Hence, $R_{\text{sum}}$ is a difference of concave functions. 
We handle this structure using SCA method. 
At each inner iteration, the second concave term is linearized using its first-order Taylor expansion around the current point $\mathbf{p}^{(t)}$, which produces a convex surrogate problem.

Furthermore, the QoS constraint $R_k \ge R_{\mathrm{th}}$ is equivalently written in SINR form as
\begin{equation}
p_k A_{k,k}
\ge
\gamma_{\mathrm{th}}
\left(
\sigma_u^2 + \sum_{j \neq k} p_j A_{k,j}
\right),
\end{equation}
where $\gamma_{\mathrm{th}} = 2^{R_{\mathrm{th}}} - 1$.
This constraint is linear in $\mathbf{p}$ and is enforced in every SCA iteration.

To guarantee feasibility, we first solve a minimum-sum-power problem under the QoS constraints to obtain a feasible starting point. If no feasible solution exists within $p_k^{\max}$, the problem is declared infeasible. The algorithm then alternates between the inner SCA updates and the outer Dinkelbach updates until convergence. 
%
\subsection{ME--RIS Phase Shift Profile ($\boldsymbol{\Phi}$)}
We optimize the ME--RIS phase shift matrix $\boldsymbol{\Phi}$ for fixed $\{\boldsymbol{v}_k, \mathbf{p}, \mathbf{T},\mathbf{U}\}$. The phase design is constrained by unit--modulus elements, which makes the problem highly non--convex. 
The phase optimization subproblem is

\begin{subequations}
\label{eq:sub_phi_original}
\begin{align}
\textup{(P1.3):} \quad
& \max_{\boldsymbol{\Phi}} \quad R_{\mathrm{sum}}(\boldsymbol{\Phi}) \nonumber \\
\mathrm{s.t.} \quad
& R_k(\boldsymbol{\Phi}) \ge R_{\mathrm{th}}, \quad \forall k \in \mathcal{K}, \label{C1_phi} \\
& \left|[\boldsymbol{\Phi}]_{n,n}\right| = 1, \quad \forall n \in \mathcal{N}. \label{C4_phi}
\end{align}
\end{subequations}

To solve ~\eqref{eq:sub_phi_original}, we use the SCA method, combined with a trust-region approach, to ensure the approximation accuracy, by letting $\boldsymbol{\vartheta}=[\vartheta_1,\ldots,\vartheta_N]^T$. At iteration $t$, we construct convex lower bounds for the SINR constraints via first--order Taylor expansions around $\boldsymbol{\vartheta}^{(t)}$, denoted by $\gamma_k^{(t)}(\boldsymbol{\vartheta})$ for $k\in\mathcal{K}$. Also, we linearize the sum rate objective to ensure stable convergence. The resulting convex surrogate problem is
\begin{subequations}
\label{eq:phi_sca}
\begin{align}
\max_{\boldsymbol{\vartheta}} \quad &
\left(\sum_{k \in \mathcal{K}} \nabla_{\boldsymbol{\vartheta}} R_{k}(\boldsymbol{\vartheta}^{(t)})\right)^{\!T}
(\boldsymbol{\vartheta} - \boldsymbol{\vartheta}^{(t)}) \label{eq:phi_sca_a} \\
\mathrm{s.t.} \quad &
\gamma_k^{(t)}(\boldsymbol{\vartheta}) \ge \gamma_{\mathrm{th}},
\quad \forall k \in \mathcal{K}, \label{eq:phi_sca_b} \\
& 0 \le \vartheta_n \le 2\pi,
\quad \forall n \in \mathcal{N}, \label{eq:phi_sca_c} \\
& \left\|\boldsymbol{\vartheta} - \boldsymbol{\vartheta}^{(t)}\right\|_2 \le \Delta^{(t)}.
\label{eq:phi_sca_d}
\end{align}
\end{subequations}
where \eqref{eq:phi_sca_d} is the trust region constraint with radius $\Delta^{(t)}$. In trust-region-based SCA, the constraint in~\eqref{eq:phi_sca_d} limits the update to a neighborhood of the current iterate to maintain the validity of the first-order approximation. The trust-region radius $\Delta^{(t)}$ is adjusted based on the accuracy of the surrogate model; it is increased when the approximation is accurate and reduced otherwise~\cite{trust1}. This approach ensures stable convergence to a stationary point~\cite{trust2,trust3}. After solving~\eqref{eq:phi_sca}, we update $\boldsymbol{\vartheta}$ and reconstruct $\boldsymbol{\Phi}=\mathrm{diag}(e^{j\vartheta_1},\ldots,e^{j\vartheta_N})$. The gradient expressions are provided in [Appendix~D, ~\cite{hokmabadi2026joint}].
\subsection{Position Optimization of MA-BS ($\mathbf{U}$) and ME-RIS ($\mathbf{T}$)}
We optimize the positions of the ME--RIS elements ($\mathbf{T}$) and the BS receive antennas ($\mathbf{U}$) while keeping all other variables fixed. These correspond to two separate subproblems. Since they have identical formulations and differ only in the optimization variable, we present them in a unified form to avoid repetition. Let $\mathbf{X}$ denote the position variable, representing either the RIS element positions $\mathbf{T}=\{\mathbf{t}_n\}_{n=1}^{N}$ or the BS antenna positions $\mathbf{U}=\{\mathbf{u}_m\}_{m=1}^{M}$. The element/antenna positions determine the field-response matrices and therefore affect the effective channels in the SINR expressions. Hence, the position optimization problem can be written as
\begin{subequations}
\label{eq:sub_pos}
\begin{align}
\textup{(P1.4):} \quad
& \max_{\mathbf{X}} \; R_{\mathrm{sum}}(\mathbf{X}) \\
\mathrm{s.t.} \quad
& R_k(\mathbf{X}) \ge R_{\mathrm{th}}, \quad \forall k \in \mathcal{K}, \\
& \mathbf{x}_i \in \mathcal{X}_i, \quad \forall i, \\
& \|\mathbf{x}_i - \mathbf{x}_j\|_2 \ge d_0, \quad \forall i \neq j.
\end{align}
\end{subequations}
where $\mathbf{X}=\mathbf{T}$ corresponds to RIS element optimization and $\mathbf{X}=\mathbf{U}$ corresponds to BS antenna optimization.

The problem is non-convex due to the nonlinear dependence of FR matrices on the position variables. We adopt a local update method based on the SCA approach in~[Appendix~C,\cite{hokmabadi2026joint}]. 

At iteration $t$, the linearized problem is
\begin{subequations}
\label{eq:sub_pos_sca}
\begin{align}
\max_{\mathbf{X}} \quad
&
\left(\sum_{k \in \mathcal{K}}
\nabla_{\mathbf{X}} R_k(\mathbf{X}^{(t)})\right)^T
\mathrm{vec}(\mathbf{X}-\mathbf{X}^{(t)}) \\
\mathrm{s.t.} \quad
& \gamma_k^{(t)}(\mathbf{X}) \ge \gamma_{\mathrm{th}},
\quad \forall k \in \mathcal{K}, \\
& \mathbf{x}_i \in \mathcal{X}_i,
\quad \forall i, \\
& \|\mathbf{x}_i - \mathbf{x}_j\|_2 \ge d_0,
\quad \forall i \neq j, \\
& \|\mathrm{vec}(\mathbf{X}-\mathbf{X}^{(t)})\|_2 \le \Delta^{(t)} .
\end{align}
\end{subequations}

This convex problem can be efficiently solved using standard solvers such as CVX. The same formulation applies to both $\mathbf{T}$ and $\mathbf{U}$ by replacing $\mathbf{X}$ with the corresponding position variable. Finally, the proposed AO algorithm is applied, solving the subproblems iteratively, as summarized in Algorithm~\ref{alg:ao}.
\begin{algorithm}[!htbp]
\small
\caption{Alternating Optimization (AO)}
\label{alg:ao}
\begin{algorithmic}[1]
\State \textbf{Define:}
$\mathcal{Z}\triangleq \{\boldsymbol{v},\boldsymbol{\Phi},\mathbf{p},\mathbf{U},\mathbf{T}\}$ as the variables set,
\State \textbf{Initialize:} feasible
$\mathcal{Z}^{(0)}
=\{\boldsymbol{v}_k^{(0)},\boldsymbol{\Phi}^{(0)},\mathbf{p}^{(0)}
\mathbf{U}^{(0)},\mathbf{T}^{(0)}\}$ randomly,
$\epsilon = 0.0001$, $n_{\max}$, and compute $\mathrm{EE}^{(0)}$.
\Repeat
    \State Update $\boldsymbol{v}_k^{(n+1)}\in\mathcal{Z}$ by solving \eqref{eq:sub_v_original} for fixed $\mathcal{Z}\setminus\{\boldsymbol{v}_k\}$.

    \State Update $\mathbf{p}^{(n+1)}\in\mathcal{Z}$ by solving \eqref{eq:sub_p_original} for fixed $\mathcal{Z}\setminus\{\mathbf{p}\}$.

    \State Update $\boldsymbol{\Phi}^{(n+1)}\in\mathcal{Z}$ by solving \eqref{eq:phi_sca} for fixed $\mathcal{Z}\setminus\{\boldsymbol{\Phi}\}$.

    \State Update $\mathbf{U}^{(n+1)}\in\mathcal{Z}$ by solving \eqref{eq:sub_pos_sca} for fixed $\mathcal{Z}\setminus\{\mathbf{U}\}$.

    \State Update $\mathbf{T}^{(n+1)}\in\mathcal{Z}$ by solving \eqref{eq:sub_pos_sca} for fixed $\mathcal{Z}\setminus\{\mathbf{T}\}$.

    \State Compute $\mathrm{EE}^{(n+1)}$ and set $n \leftarrow n+1$.
\Until{$\mathrm{EE}^{(n)}-\mathrm{EE}^{(n-1)} \le \epsilon$ or $n \ge n_{\max}$}
\State \textbf{Output:} $\mathcal{Z}^\star = \mathcal{Z}^{(n)}$.
\end{algorithmic}
\end{algorithm}
\subsection{Computational Complexity Analysis}
We analyze the per-iteration complexity of the proposed AO algorithm, where $I_a$ denotes the number of AO iterations. 

The uplink postcoding vectors are obtained via generalized eigenvalue decomposition with complexity $\mathcal{O}(M^3)$~\cite{golub}. The power allocation problem is solved using Dinkelbach's method with SCA, where each SCA iteration requires SINR evaluations with complexity $\mathcal{O}(K^2 M)$~\cite{kumar2022novel}, repeated for $I_s^p$ inner iterations and $I_d$ Dinkelbach updates~\cite{dinkelbach}, giving $\mathcal{O}(I_d I_s^p K^2 M)$.
The RIS phase optimization uses SCA with $I_s^\phi$ iterations, each solving a convex quadratic program with complexity $\mathcal{O}(N^{3.5})$~\cite{boyd}. The position updates for the BS antennas and RIS elements require $\mathcal{O}(I_s^u M N L)$ and $\mathcal{O}(I_s^t N^2 L)$ operations, respectively, where $L=\max\{L_{\text{RB}}, L_{\text{Bu}}, L_{\text{Ru}}\}$. The overall per-iteration complexity is
\begin{equation}
\mathcal{O}\!\left(
M^3
+ I_d I_s^p K^2 M
+ I_s^\phi N^{3.5}
+ (I_s^u M + I_s^t N) N L
\right).
\end{equation}
When $N$ is large, the RIS phase and position updates dominate. The fixed-position benchmark reduces complexity by removing position optimization terms, at the cost of lower EE performance.

\section{Numerical Results} \label{sec:results}
We evaluate the performance of the proposed AO-based algorithm via Monte Carlo simulations. The results are averaged over $100$ independent channel realizations. For comparison, we consider four setups based on the mobility of the BS antennas and RIS elements; movable/fixed antenna (MA/FA) and movable/fixed RIS element (ME/FE). The considered scenarios are: 1) \emph{MA--ME}, 2) \emph{FA--ME}, 3) \emph{MA--FE}, and 4) \emph{FA--FE}. Unless otherwise stated, all schemes are evaluated under the setup of $N=49, K=4, M=8, L=4$, $R_{\text{th}} = 1.5$~bps/Hz and maximum transmit power of $P_{\max} = 20$~dBm, $\forall k  \in \mathcal{K}$.

The BS is placed at the origin $(0, 0, 15)m$ and the RIS at $(10, 10, 10)m$. The $K$ users with a height of $1.5m$ are randomly distributed within a circular area around the RIS at distances of $50-70m$. The movable region for each BS antenna and RIS element is a square of side $A = 4\lambda$, where 
$\lambda$ denotes the carrier wavelength at $f_c = 3$~GHz. The minimum inter-element spacing is set to $d_0 = \lambda/2$. The rest of the simulation parameters are listed in Table~\ref{tab:params}.
\begin{table}[!htbp]
\caption{Simulation Parameters}
\begin{center}
\begin{tabular}{|l|c|}
\hline
\textbf{Parameter} & \textbf{Value} \\
\hline
Path loss exponent for $\{\mathbf{h}_k, \mathbf{H}, \mathbf{g}_k\}$ & $\alpha = \{3.9, 2, 2.2\}$ \\
\hline
Noise power ($\sigma^2$) & $-90$~dBm \\
\hline
Circuit power consumption ($P_c$) & $20$~dBm \\
\hline
Power amplifier efficiency ($\eta$)& $0.3$ \\
\hline
\end{tabular}
\label{tab:params}
\end{center}
\end{table}

\begin{figure}[!t]
    \centering
    \includegraphics[width=0.7\linewidth]{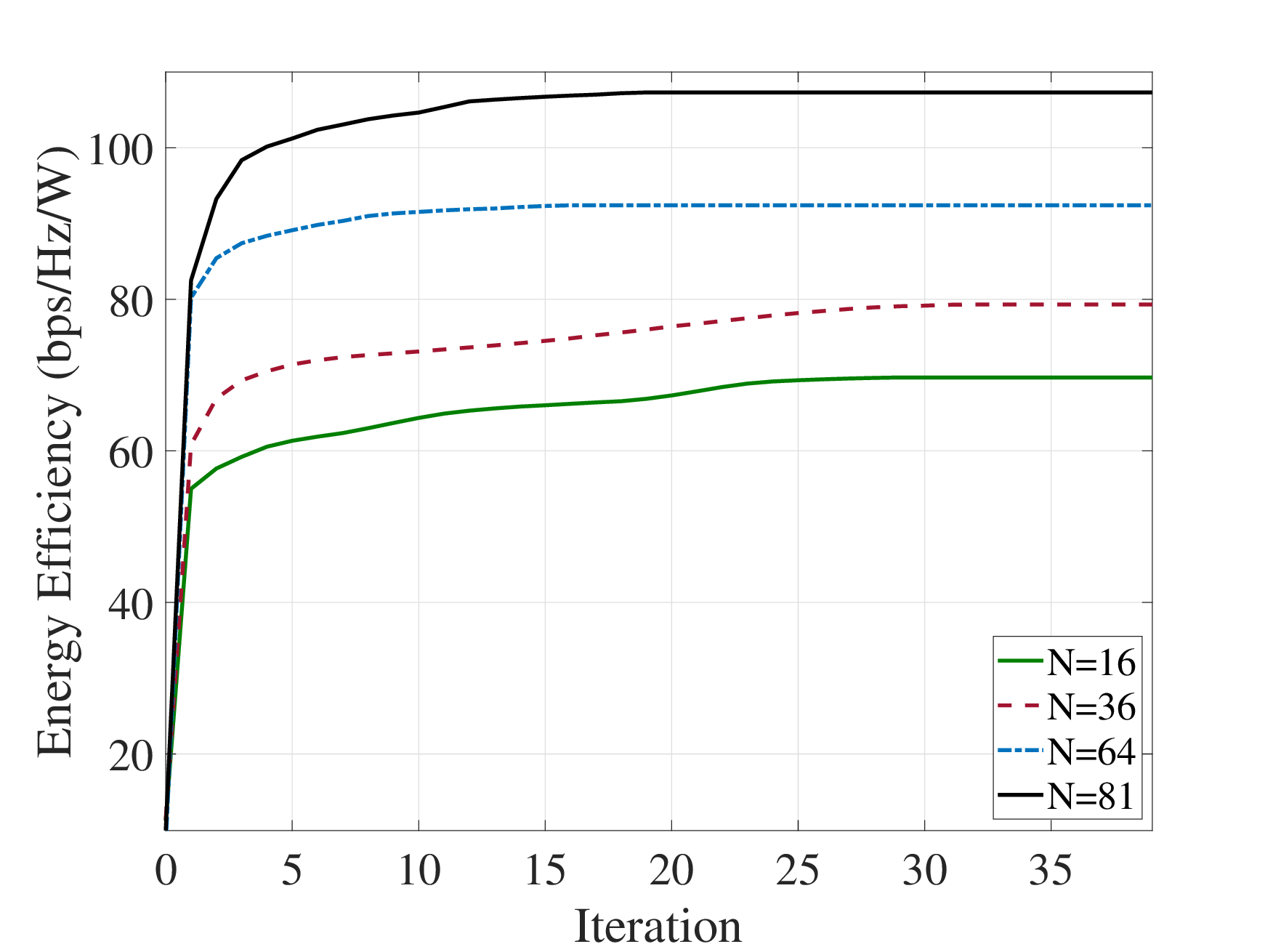}
    \caption{Convergence behavior of the AO algorithm}
    \label{fig:convergence}
\end{figure}

Fig.~(\ref{fig:convergence}) shows the convergence behavior of Algorithm~1 for the \emph{MA--ME} scheme for different $N$ values. The proposed AO algorithm converges within approximately 20 iterations in all cases, confirming its efficiency and stability.

\begin{figure}[!t]
\centering
\includegraphics[width=0.7\linewidth]{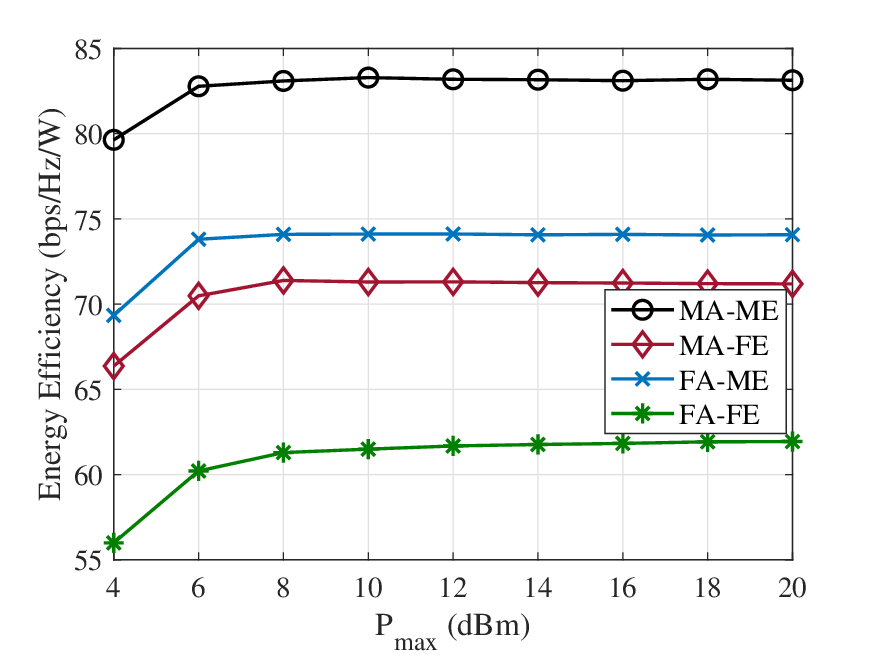}
\caption{EE performance versus $P_{\max}$.}
\label{fig:vs_P}
\end{figure}

Fig.~(\ref{fig:vs_P}) illustrates the system EE versus the maximum user transmit power, $P_{\max}$, for the minimum achievable rate of $R_{\text{th}} = 0.5$~bps/Hz. The EE increases with $P_{\max}$ at low power levels since higher transmit power improves the achievable rates while the power consumption remains relatively small. However, beyond about 6--8 dBm, all curves gradually saturate because the rate grows logarithmically while the power consumption increases linearly, resulting in diminishing EE gains. The proposed \emph{MA--ME} scheme achieves the highest EE across the entire power range, outperforming the \emph{FA--FE} baseline by about $42\%$ at $P_{\max} = 4$ dBm and $34\%$ at higher power levels. The larger gap at low power highlights the benefit of jointly optimizing the BS antenna and RIS element positions when transmit power is limited. Among the partially movable setups, \emph{FA--ME} consistently outperforms \emph{MA--FE}, indicating that RIS element mobility provides greater EE gains than BS antenna mobility in this setup. Finally, the \emph{FA--FE} baseline achieves the lowest EE, highlighting that fixed-position setups cannot fully exploit the available spatial degrees of freedom, resulting in limited beamforming and propagation control.

\begin{figure*}[!t]
\centering
\begin{subfigure}[t]{0.32\linewidth}
    \centering
    \includegraphics[width=\linewidth]{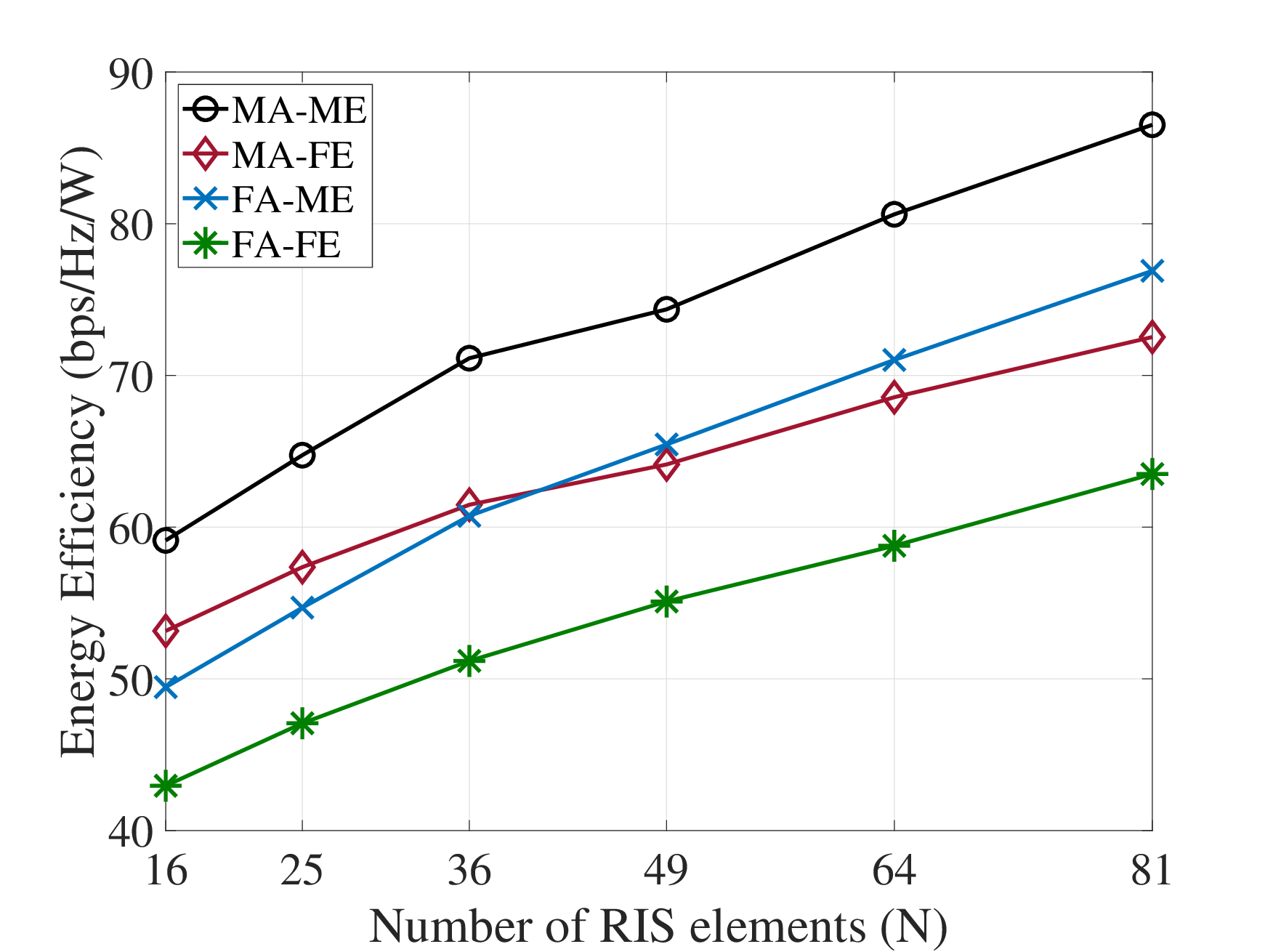}
    \caption{}
    \label{fig:vs_N}
\end{subfigure}
\begin{subfigure}[t]{0.32\linewidth}
    \centering
    \includegraphics[width=\linewidth]{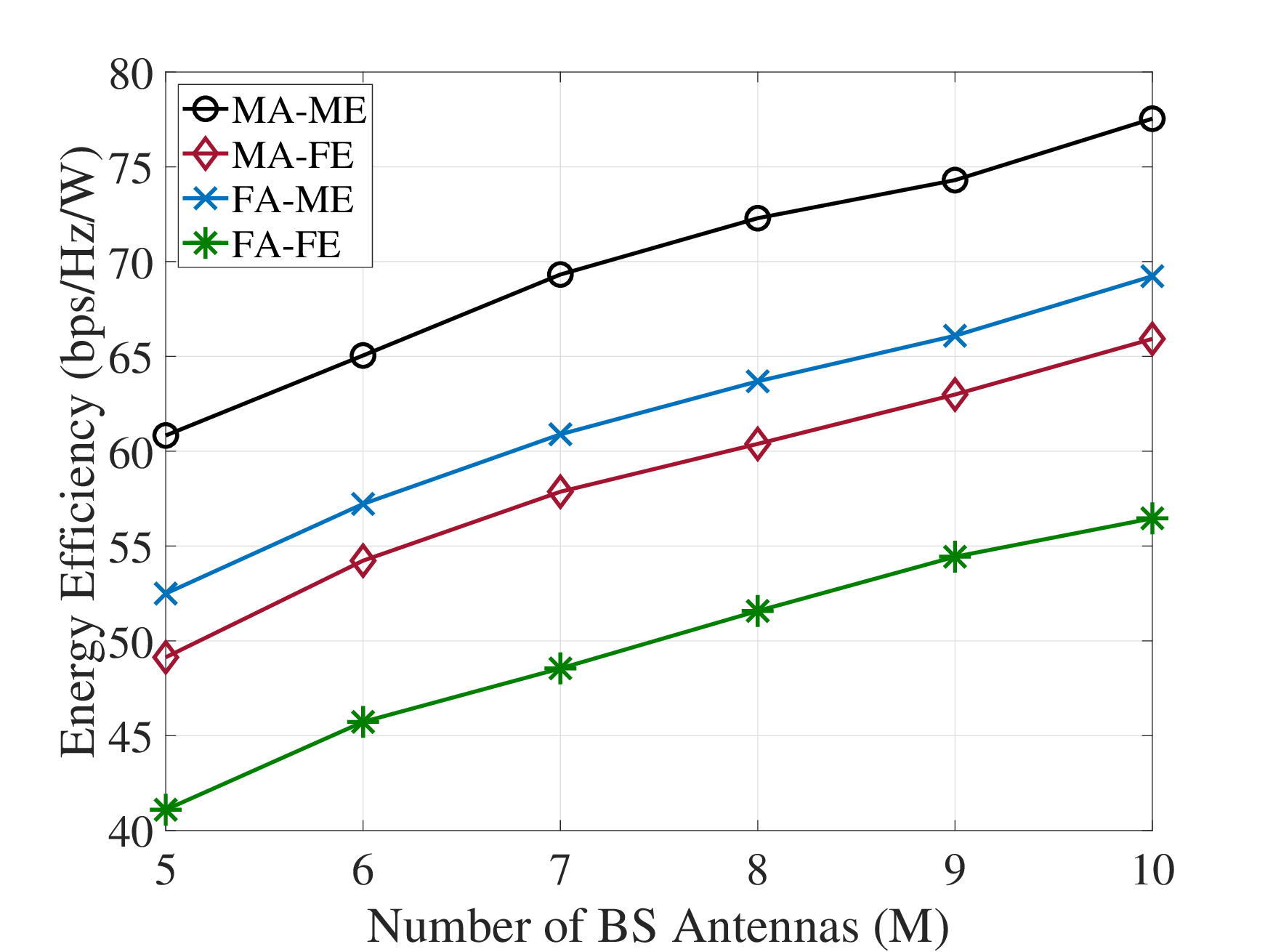}
    \caption{}
    \label{fig:vs_M}
\end{subfigure}
\begin{subfigure}[t]{0.32\linewidth}
    \centering
    \includegraphics[width=\linewidth]{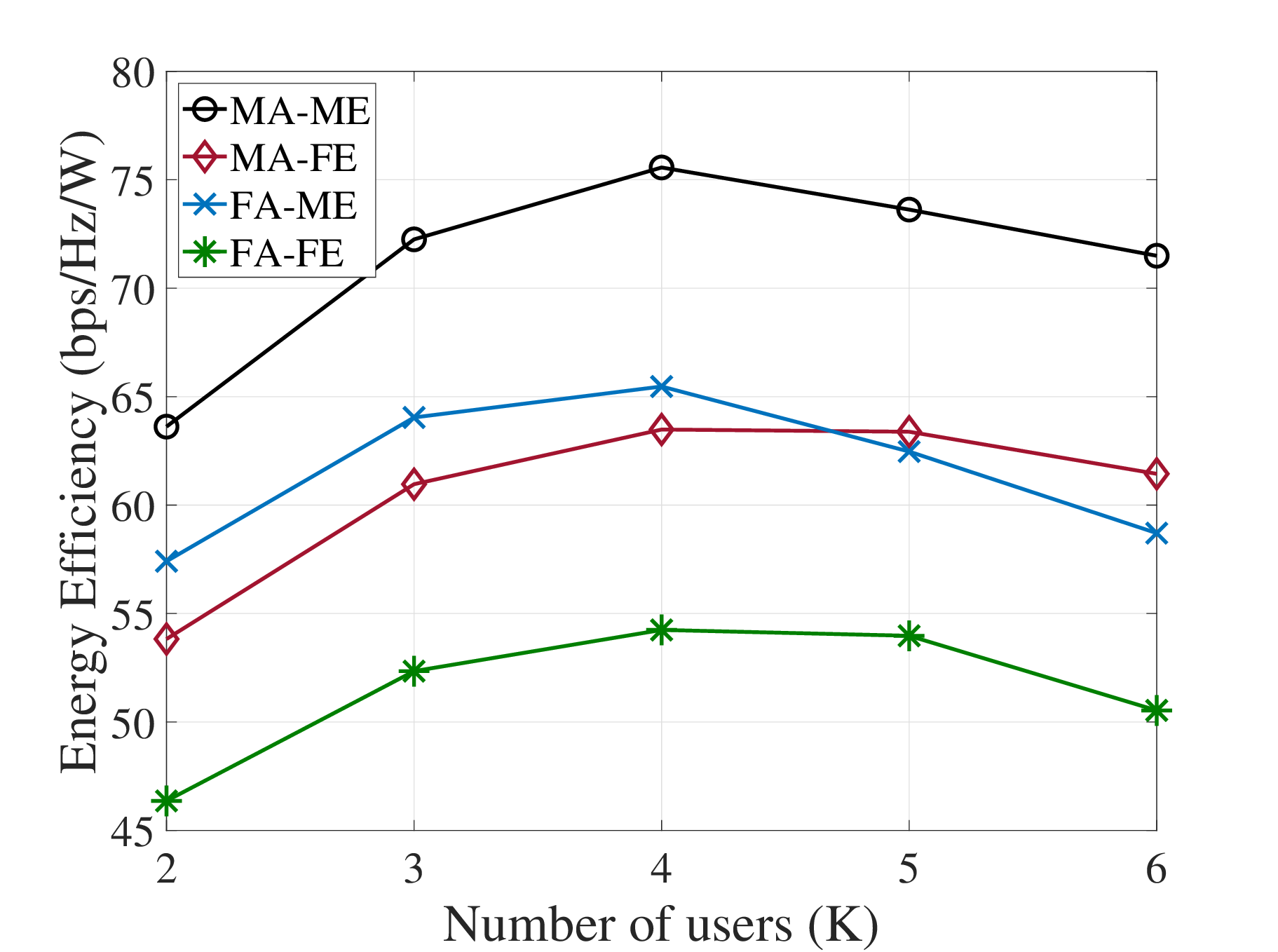
    }
    \caption{}
    \label{fig:vs_K}
\end{subfigure}
\caption{EE performance versus $N$, $M$, and $K$.}
\label{fig:main1}
\vspace{-0.3cm}
\end{figure*}

Fig.~(\ref{fig:vs_N}) presents the system EE as a function of the number of RIS elements, $N$. The EE increases with $N$ for all schemes because a larger RIS provides greater passive beamforming gain and extended coverage, improving the sum rates without a proportional increase in power consumption. The proposed \emph{MA--ME} scheme consistently achieves the highest EE across all values of $N$, outperforming the \emph{FA--FE} baseline by approximately 35\% at $N=49$. This gain results from the joint mobility of the BS antennas and RIS elements, enabling better receive beamforming and reflected link alignment than fixed or partially movable scenarios. Furthermore, the \emph{MA--FE} outperforms \emph{FA--ME} at small values of $N$ since the RIS contribution is limited, whereas the trend reverses as $N$ increases and the RIS becomes more dominant. Finally, the \emph{FA--FE} baseline achieves the lowest EE in all cases, as discussed previously.

Fig.~(\ref{fig:vs_M}) shows the system EE for different numbers of BS receive antennas, $M$. The EE increases with $M$ for all schemes, since more antennas provide stronger receive beamforming gain and expand the spatial degrees of freedom available at the BS, enabling sharper beamforming toward each user and better interference suppression across users. The proposed \emph{MA--ME} scheme consistently achieves the highest EE, outperforming the \emph{FA--FE} baseline by about $37\%$ at $M=10$, indicating that antenna position optimization becomes more effective as the available spatial degrees of freedom increase. 
Unlike the previous case, no crossover occurs between \emph{MA--FE} and \emph{FA--ME}; \emph{FA--ME} consistently performs better, showing that for a fixed RIS size $N=49$, repositioning RIS elements provides larger EE gains than moving BS antennas. Also, the \emph{FA--FE} baseline achieves the lowest EE, consistent with the observations discussed above.

Fig.~(\ref{fig:vs_K}) presents the EE of the system for different numbers of uplink users, $K$. As $K$ increases, the EE first rises due to multiuser diversity gain, where additional users contribute to the overall sum rate. 
However, beyond a certain point, inter-user interference grows faster than the sum rate improvement, causing the EE to peak and then slightly decrease at larger $K$. The proposed \emph{MA--ME} scheme consistently achieves the highest EE across all values of $K$, outperforming the \emph{FA--FE} baseline by approximately $39\%$ at $K=5$.

\section{Conclusion}
\label{sec:conclusion}
In this paper, we studied an energy efficiency maximization problem for an uplink multiuser system with an ME-RIS and a MA-BS. We jointly optimized the postcoder vectors, user transmit powers, RIS phase 
shifts, and antenna/element positions using an AO-based algorithm that combines Dinkelbach's method and SCA. Simulation results confirmed that the proposed scheme consistently outperforms fixed-position benchmarks, as the additional spatial degrees of freedom from position optimization improve channel conditions and suppress inter-user interference with lower power consumption.

\FloatBarrier
\bibliographystyle{IEEEtran}
\bibliography{references.bib}

@article{basar2019wireless,
  title={{Wireless Communications Through Reconfigurable Intelligent Surfaces}},
  author={Basar, Ertugrul and Di Renzo, Marco and De Rosny, Julien and Debbah, Merouane and Alouini, Mohamed-Slim and Zhang, Rui},
  journal={{IEEE Access}},
  volume={7},
  pages={116753--116773},
  year={2019},
  publisher={IEEE}
}

@article{huang2019ris,
  title={{Reconfigurable Intelligent Surfaces for Energy Efficiency in Wireless Communication}},
  author={Huang, Chongwen and Zappone, Alessio and Alexandropoulos, George C and Debbah, M{\'e}rouane and Yuen, Chau},
  journal={{IEEE Transactions on Wireless Communications}},
  volume={18},
  number={8},
  pages={4157--4170},
  year={2019},
  publisher={IEEE}
}

@article{wu2019irs,
  title={{Intelligent Reflecting Surface Enhanced Wireless Network via Joint Active and Passive Beamforming}},
  author={Wu, Qingqing and Zhang, Rui},
  journal={{IEEE Transactions on Wireless Communications}},
  volume={18},
  number={11},
  pages={5394--5409},
  year={2019},
  publisher={IEEE}
}

@article{zhu2024ma_mag,
  title={{Movable Antennas for Wireless Communication: Opportunities and Challenges}},
  author={Zhu, Lipeng and Ma, Wenyan and Zhang, Rui},
  journal={{IEEE Communications Magazine}},
  volume={62},
  number={6},
  pages={114--120},
  year={2023},
  publisher={IEEE}
}

@article{zhu2024ma_twc,
  title={{Movable-Antenna Enhanced Multiuser Communication via Antenna Position Optimization}},
  author={Zhu, Lipeng and Ma, Wenyan and Ning, Boyu and Zhang, Rui},
  journal={{IEEE Transactions on Wireless Communications}},
  year={2023},
  publisher={IEEE}
}

@article{zhang2025ris_ma,
  title={{Sum-Rate Enhancement for RIS-Assisted Movable Antenna Systems: Joint Transmit Beamforming, Reflecting Design, and Antenna Positioning}},
  author={Zhang, Beihua and Xu, Kui and Xia, Xiaochen and Hu, Guojie and Wei, Chen and Li, Chunguo and Cheng, Kaixin},
  journal={{IEEE Transactions on Vehicular Technology}},
  volume={74},
  number={3},
  pages={4376--4392},
  year={2024},
  publisher={IEEE}
}

@article{wei2024ma_irs,
  title={{Joint Beamforming and Antenna Position Optimization for Movable Antenna-Assisted Spectrum Sharing}},
  author={Wei, Xin and Mei, Weidong and Wang, Dong and Ning, Boyu and Chen, Zhi},
  journal={{IEEE Wireless Communications Letters}},
  year={2024},
  publisher={IEEE}
}

@article{hokmabadi2026joint,
  title={{Joint Beamforming and Position Optimization for Movable-Antenna and Movable-Element RIS-Aided Full-Duplex 6G MISO Systems}},
  author={Hokmabadi, Ayda Nodel and Assi, Chadi},
  journal={{arXiv Preprint arXiv:2601.08922}},
  year={2026}
}

@article{zhao2026exploiting,
  title={{Exploiting Movable-Element STARS for Wireless Communications}},
  author={Zhao, Jingjing and Zhou, Quan and Mu, Xidong and Cai, Kaiquan and Zhu, Yanbo and Liu, Yuanwei},
  journal={{IEEE Transactions on Wireless Communications}},
  year={2026},
  publisher={IEEE}
}

@article{amhaz2026meta,
  title={{Meta-Learning-Driven Resource Optimization in Full-Duplex ISAC With Movable Antennas}},
  author={Amhaz, Ali and Khisa, Shreya and Elhattab, Mohamed and Assi, Chadi and Sharafeddine, Sanaa},
  journal={{IEEE Wireless Communications Letters}},
  year={2026},
  publisher={IEEE}
}

@article{MERIS1,
  title={{Movable-Element STARS-Aided Secure Communications}},
  author={Zhao, Jingjing and Xu, Qian and Cai, Kaiquan and Zhu, Yanbo and Mu, Xidong and Liu, Yuanwei},
  journal={{IEEE Transactions on Vehicular Technology}},
  year={2025},
  publisher={IEEE}
}

@article{MERIS2,
  title={{Joint Beamforming and Antenna Position Optimization for IRS-Aided Multi-User Movable Antenna Systems}},
  author={Geng, Yue and Cheng, Tee Hiang and Zhong, Kai and Teh, Kah Chan and Wu, Qingqing},
  journal={{IEEE Transactions on Wireless Communications}},
  year={2025},
  publisher={IEEE}
}

@article{MERIS3,
  title={{Element Coefficient and Position Optimization for Flexible Intelligent Metasurface-Aided Movable Antenna Communications}},
  author={Bu, Qifei and Yang, Songjie and Ge, Jianhua and Ning, Boyu and Yuen, Chau},
  journal={{IEEE Wireless Communications Letters}},
  year={2025},
  publisher={IEEE}
}

@article{zhang2024ma_ris_geometry,
  title={{RIS-Aided Wireless Communication with Movable Elements: Geometry Impact on Performance}},
  author={Zhang, Yan and Dey, Ipsita and Marchetti, Nicola},
  journal={{arXiv Preprint arXiv:2405.00141}},
  year={2024}
}

@inproceedings{wei2024movable_ris,
  title={{Movable Antennas Meet Intelligent Reflecting Surface: When Do We Need Movable Antennas?}},
  author={Wei, Xin and Mei, Weidong and Wu, Qingqing and Ning, Boyu and Chen, Zhi},
  booktitle={{2025 IEEE Wireless Communications and Networking Conference (WCNC)}},
  pages={1--6},
  year={2025},
  organization={IEEE}
}

@article{li2024mis,
  title={{Movable Intelligent Surface (MIS) for Wireless Communications: Architecture, Modeling, Algorithm, and Prototyping}},
  author={Zheng, Ziyuan and Wu, Qingqing and Chen, Wen and Wu, Xiangming and Zhu, Weiren},
  journal={{IEEE Transactions on Wireless Communications}},
  year={2025},
  publisher={IEEE}
}

@article{zhao2025me_ris,
  title={{Movable-Element RIS-Aided Wireless Communications: An Element-Wise Position Optimization Approach}},
  author={Zhao, Jingjing and Huang, Qingyi and Cai, Kaiquan and Zhou, Quan and Mu, Xidong and Liu, Yuanwei},
  journal={{IEEE Communications Letters}},
  year={2026},
  publisher={IEEE}
}

@article{rayleigh1,
  title={{Rayleigh Quotient Based Optimization Methods for Eigenvalue Problems}},
  author={Li, Ren-Cang},
  journal={{Matrix Functions and Matrix Equations}},
  year={2015},
  publisher={World Scientific}
}

@inproceedings{rayleigh3,
  title={{Secure Full-Duplex Communication via Movable Antennas}},
  author={Ding, Jingze and Zhou, Zijian and Wang, Chenbo and Li, Wenyao and Lin, Lifeng and Jiao, Bingli},
  booktitle={{IEEE Global Communications Conference}},
  pages={885--890},
  year={2024},
  organization={IEEE}
}

@article{dinkelbach,
  author={W. Dinkelbach},
  title={{On Nonlinear Fractional Programming}},
  journal={{Management Science}},
  volume={13},
  number={7},
  pages={492--498},
  year={1967},
  month=mar
}

@inproceedings{trust1,
  title={{Networked ISAC for Low-Altitude Economy: Transmit Beamforming and UAV Trajectory Design}},
  author={Cheng, Gaoyuan and Song, Xianxin and Lyu, Zhonghao and Xu, Jie},
  booktitle={{2024 IEEE/CIC International Conference on Communications in China (ICCC)}},
  pages={78--83},
  year={2024},
  organization={IEEE}
}

@inproceedings{trust2,
  title={{Joint Trajectory and Beamforming Design for UAV-Enabled Integrated Sensing and Communication}},
  author={Lyu, Zhonghao and Zhu, Guangxu and Xu, Jie},
  booktitle={{ICC 2022 - IEEE International Conference on Communications}},
  year={2022},
  organization={IEEE}
}

@article{trust3,
  title={{Dual-UAV-Aided Covert Communications for Air-to-Ground ISAC Networks}},
  author={Sun, Jingke and Yang, Liang and Boulogeorgos, Alexandros-Apostolos A and Tsiftsis, Theodoros A and Liu, Hongwu},
  journal={{arXiv Preprint arXiv:2506.00601}},
  year={2025}
}

@book{golub,
  title={{Matrix Computations}},
  author={Golub, Gene H and Van Loan, Charles F},
  year={2013},
  publisher={JHU Press}
}

@book{boyd,
  title={{Convex Optimization}},
  author={Boyd, Stephen and Vandenberghe, Lieven},
  year={2004},
  publisher={Cambridge University Press}
}

@article{kumar2022novel,
  title={{A Novel SCA-Based Method for Beamforming Optimization in IRS/RIS-Assisted MU-MISO Downlink}},
  author={Kumar, Vaibhav and Zhang, Rui and Di Renzo, Marco and Tran, Le-Nam},
  journal={{IEEE Wireless Communications Letters}},
  volume={12},
  number={2},
  pages={297--301},
  year={2023},
  publisher={IEEE}
}
\end{document}